\documentclass[preprint,amsmath,amssymb,aps,pra,floatfix,nofootinbib,superscriptaddress]{revtex4-1}
\usepackage{amsfonts}
\usepackage{ifpdf}
\usepackage{mathrsfs}
\usepackage{graphicx}
\usepackage{dcolumn}
\usepackage{bm}
\begin{document}
\title{Quench Dynamics of Entanglement in an Opened Anisotropic Spin-1/2 Heisenberg Chain}
\date{\today}
\author{Jie Ren}
\affiliation{Department of Physics and Jiangsu Laboratory of
Advanced Functional Materials, Changshu Institute of
Technology, Changshu, Jiangsu 215500, People's Republic of China}

\affiliation{School of Physical Science and Technology, Suzhou University,
Suzhou, Jiangsu 215006, People's Republic of China}
\author{Shiqun Zhu}

\affiliation{School of Physical Science and Technology, Suzhou University,
Suzhou, Jiangsu 215006, People's Republic of China}

\begin{abstract}
The quantum entanglement dynamics of a one-dimensional spin-1/2
anisotropic XXZ model is studied using the method of the adaptive
time-dependent density-matrix renormalization-group when two cases
of quenches are performed in the system. An anisotropic interaction
quench and the maximum number of domain walls of staggered
magnetization quench are considered. The dynamics of pairwise
entanglement between the nearest two qubits in the spin chain is
investigated. The entanglement of the two-spin qubits can be created
and oscillates in both cases of the quench. The anisotropic
interaction has a strong influence on the oscillation frequency of
entanglement.

\vskip 0.6 cm



\end{abstract}

\maketitle

\section{Introduction}
Entanglement generation and distribution is an important problem in
performing quantum-information tasks, such as quantum computation
and quantum teleportation \cite{Nielsen,Bennett,Wootters}. Many
results showed that entanglement existed naturally in the spin chain
when the temperature is at zero. Due to its potential applications,
the pairwise entanglement of anisotropic Heisenberg model was
extensively studied recently \cite{Wang}. A typical example is the
exactly solved one-dimensional infinite-lattice anisotropic $XY$
model using the Jordan-Wigner transform to obtain the pairwise
entanglement \cite{Osborne,Osterloh,Amico,Amicoa}. By using the
Bethe ansatz solution, the entanglement in infinite isotropic
Heisenberg rings was calculated \cite{Poulsen}. In recent years, the
study of the dynamics of entanglement has attracted much attention
due to the manipulation of quantum systems. The dynamics of
entanglement in different systems were investigated \cite{Sengupta,
Calabrese,Bravyi,Eisert,Cincio,Fagotti,Li,Furman,Chiara, Zhang}. The
effects of anisotropy interaction on the evolution of entanglement
were investigated \cite{Chiara}, such a global quench could actually
be realized for atoms in optical lattices \cite{Greiner}. Moreover,
the effects of a bond defect on the entanglement dynamics in a local
quench were studied \cite{Eisler} and the evolution of
one-dimensional quantum lattice systems was investigated when the
ground state was perturbed by altering one site in the middle of the
chain \cite{Tonya,Tonyb,Santosa,Santosb,Santosc,Perales}.
Recently, Bose exploited a global quench dynamics in spin chains for
distant pairwise entanglement, which could be used for quantum
communication \cite{Bose}. It would be interesting to investigate
the overall pairwise entanglement dynamics between the nearest two
qubits in the spin chain when several typical quenches are performed
in the system.

In this paper, quench dynamics of entanglement in an opened
anisotropic Heisenberg spin chain is analyzed. In section II, the
Hamiltonian of an opened anisotropic Heisenberg spin chain is
presented. In section III, the dynamics of pairwise entanglement
between the nearest two qubits in the spin chain is studied when two
kinds of typical quench are performed in the system. In section IV,
a discussion concludes the paper.

\section{Hamiltonian of the System}

The Hamiltonian of an opened Heisenberg XXZ chain with N sites is
given by

\begin{equation}
\label{eq1}H=\sum_{i=1}^{N-1}[J_{xy}(\sigma^x_{i}\sigma^x_{i+1}
+\sigma^y_{i}\sigma^y_{i+1})+J_z\sigma^z_{i}\sigma^z_{i+1}],\\
\end{equation}
where $\sigma^{\alpha}_i(\alpha=x, y, z)$ are Pauli operators on the
$i$-th site, $N$ is the length of the spin chain, $J_{xy}, J_z$
denotes the couplings in the $xy$-plane and the $z$-axis
respectively. For simplicity, the couplings in the plane $J_{xy}=1$
is considered. An opened boundary condition (OBC) is assumed because
the antiferromagnetic Heisenberg spin chain with OBC can be achieved
artificially in the experiment \cite{hirjibehedin}.

In the paper, the concurrence is chosen as a measurement of the
pairwise entanglement \cite{Wootters}. The concurrence $C$ is
defined as

\begin{equation}
\label{eq2} C = \max \{{\lambda_1 - \lambda_2 - \lambda_3 -
\lambda_4 ,0}\},
\end{equation}
where the quantities $\lambda_i (i=1, 2, 3, 4)$ are the square roots
of the eigenvalues of the operator $\varrho = \rho_{12}(\sigma_1^y
\otimes \sigma_2^y)\rho_{12}^\ast (\sigma_1^y \otimes \sigma_2^y)$.
They are in descending order. The case of $C=1$ corresponds to the
maximum entanglement between the two qubits, while $C=0$ means that
there is no entanglement between the two qubits.

\section{Pairwise Entanglement Dynamics}

In this section, the entanglement dynamics of a spin$-1/2$
antiferromagnetic Heisenberg chain is analyzed when two kinds of
quenches are performed in the system. One kind of quench is
performed by abruptly varying the anisotropic interaction from very
large value down to finite values. Another kind of quench is
performed by sudden release of the domain walls of staggered
magnetization. It is known that it is hard to calculate the dynamics
of entanglement because of the lack of knowledge of eigenvalues and
eigenvectors of the Hamiltonian. For models that are not exactly
solvable, most of researchers resort to exact diagonalization to
obtain the ground state for small system size. While for small
system size, some phenomena might be missing. If the system size is
large, other kinds of phenomena can appear and cen be seen much
clearer.

For large system size, the adaptive time-dependent density-matrix
renormalization-group (t-DMRG) can be applied with a second order
Trotter expansion of Hamiltonian as described in
\cite{White,Vidalb}. In order to check the accuracy of the results
of t-DMRG, the results of exact diagonalization can be considered as
a benchmark for a small size system. In the simulation, a Trotter
slicing $\delta t=2.5\cdot10^{-2}$ and Matlab codes of t-DMRG with
double precision are performed with a truncated Hilbert space of
$m=60$. It turns out that a typically discarded weight of $\delta
\rho = 10^{-8}$ can keep the relative error $\delta C$ in $C$ below
$10^{-6}$ for a chain of $L = 60$ sites with time $t\leq40/J$.

\subsection{Anisotropic Interaction Quench}

To start t-DMRG, some initial states need to be prepared. Firstly,
the initial state can be prepared to be a perfect antiferromagnetic
N\'{e}el state. The perfect antiferromagnetic N\'{e}el state is the
ground state of Eq. (1) with $J_z\rightarrow+\infty$. It is given by

\begin{equation}
\label{eq3} |ini\rangle_1=|\uparrow\downarrow\uparrow\downarrow
\cdots \uparrow\downarrow\rangle.
\end{equation}
Such a state has been achieved with high fidelity by using decoupled
double wells \cite{Trotzky}. The procedure adopted for the initial
state $|ini\rangle_1$ is to calculate it as the ground state of a
suitably chosen Hamiltonian $H_{ini}=H(J_z\rightarrow+\infty)$. The
quantum quench is performed when the anisotropic interaction
parameter is abruptly varied from an initially very large value down
to finite values \cite{Bose}.

The entanglement $C_{i,i+1}$ between the nearest two qubits is
plotted in Fig. 1 as a function of the spin site $i$ and the time
$t$ when $J_z=1.0$. The three dimensional entanglement $C_{i,i+1}$
is plot in Fig. 1(a). The contour line is plotted in Fig. 1(b). From
Fig. 1, it is clear that there is an entanglement wall for small
time $t$. It seems that the sudden change of $J_z$ from extremely
large value ($J_z\rightarrow +\infty$) to a finite value creates the
entanglement wall \cite{Amico,Amicoa}. The entanglement in the
range of $10<i<50$ is quite similar. When the time $t=0$, the
entanglement is zero. When the time increases, the entanglement
creates and increases from zero. The entanglement reaches a peak at
the time $t=0.7$. When the time increases further, the entanglement
value maintains in the extent $[0,0.1]$. In Fig. 1, it could also be
seen that when $i<10$ and $50>i$, the entanglement oscillates
acutely. The maximal value of entanglement is higher than the
maximal value in the extent of $10<i<50$. This is due to the effects
of the open boundary condition \cite{Chiara,Wanga}. From Fig. 1, it
is clear that the entanglement $C_{i, i+1}$ is symmetric about
$i=30$. This is mainly due to the fact that several sources in a
lattice emit oppositely moving pairs of entangled quasi-particles
\cite{Amicoa}.

In order to avoid the boundary effect, the entanglement $C_{30,31}$
of the central two qubits is plotted as a function of the time $t$
for different anisotropic interaction in Fig. 2. The anisotropic
interaction has little influence on the height of the first peak of
the entanglement. When $J_z=0, 0.5$, the entanglement drops down to
zero after reaching the peak. The entanglement cannot generate at
any longer time. When $J_z=1.0$, the entanglement can appear and
disappear alternatively. When $J_z=1.5, 2.0$, the entanglement can
create, oscillate and does not disappear for longer time. Except the
first peak in $C_{30,31}$, the height of oscillation of entanglement
with $J_z=2.0$ is higher than that with $J_z=1.0, 1.5$. This is due
to the strong dependence of the damping coefficient of the
entanglement wave on the anisotropic parameter. When $t$ increases,
the relative height of the first peak and the average value of
$C_{30, 31}$ becomes smaller with increasing $J_z$. The increase of
$J_z$ reduces the relative height in $C_{30, 31}$
\cite{Amico,Amicoa}.

\subsection{Domain Walls of Staggered Magnetization Quench}

Secondly, the initial state can be prepared to be an inhomogeneous
initial state. It can be given by \cite{Gobert,Lea}

\begin{equation}
\label{eq4} |ini\rangle_2
=|\uparrow_1\cdots\uparrow_{N/2}\downarrow_{N/2+1}\cdots\downarrow_{N}\rangle,
\end{equation}
where all spins on the left half are pointing up along the z axis,
while all spins on the right half pointing down. The state contains
many high-energy excitations and is thus far from equilibrium. It
can be considered as a state with almost the maximum number of
domain walls of staggered magnetization. The quantum quench is
performed when the domain walls are suddenly released in a
Heisenberg XXZ chain with different anisotropic interaction
\cite{Gobert}.

The entanglement $C_{i,i+1}$ between the nearest two qubits is
plotted as a function of spin site $i$ and time $t$ in Fig. 3 when
$J_z=0$. The three dimensional plot of $C_{i,i+1}$ is shown in Fig.
3(a). The counter line of $C_{i,i+1}$ is plotted in Fig. 3(b). From
Fig. 3, it is seen that the entanglement $C_{i,i+1}$ can be
generated. The entanglement is mirror symmetric about the line of
$i=30$. The entanglement $C_{30,31}$ generates firstly, then the
entanglement $C_{29,30}$ and $C_{31,32}$ generate, and so on. The
period of entanglement $C_{i,i+1}$ generation is a linear function
of $|i-30|$. The notion of diffusive dynamics of pairwise
entanglement in the system can be clearly seen \cite{Gobert,langer}.
The two central neighboring sites are in the state
$|\uparrow_{N/2}\downarrow_{N/2+1}\rangle$. Due to the exchange
interaction, they will be entangled after the time $t=0$. The
entanglement, initially localized on the two central neighboring
sites of the chain will spread owing to the exchange interaction
\cite{Amico,Amicoa}.

In order to see the effects of the anisotropic interaction, the
entanglement $C_{30,31}$ and the entanglement $C_{25,26}$ of the
central two qubits is plotted in Fig. 4 as a function of the time
$t$ for different anisotropic interaction. In Fig. 4(a), the
entanglement $C_{30,31}$ generates immediately after the time $t=0$.
The entanglement reaches the first peak rapidly and then oscillates
around a constant. The anisotropic interaction has a weak influence
on the first peak of the entanglement. The first peak decreases
slightly when the anisotropic interaction increases. When $J_z=0.0$,
the entanglement oscillates around $0.35$. When $J_z=0.5$, the
entanglement oscillates around $0.12$. When $J_z=1.5,2.0$, the
entanglement oscillates around $0.40$ and exists all the time. For
the above cases, the entanglement maintains for a long time. While
for $J_z=1$, the entanglement appears and disappears alternatively.
After $t=37$, the entanglement disappears all the time. It is clear
that the entanglement $C_{30, 31}$ of the central two qubits
decreases as the anisotropic interaction $J_z$ increases and then
increases as $J_z$ increases further and finally almost saturates.
In Fig. 4(b), the entanglement $C_{25,26}$ generates after a short
time. This time delay is caused by the finite propagation speed of
the entanglement from the central pair to side pairs. The
entanglement reaches the first peak rapidly and then oscillates
around a constant. The anisotropic interaction has a strong
influence on the first peak of the entanglement. The height of the
first peak decreases when the anisotropic interaction increases.
When $J_z=0$, the first peak is $C_{25,26}^{peak(1)}=0.427$, while
the peak is $C_{25,26}^{peak(1)}=0.259$ when $J_z=0.5$. The
anisotropic interaction has a feeble influence on the period when
the entanglement reaches the first peak. The time of the
entanglement reaching the first peak decreases slightly when the
anisotropic interaction increases. After the entanglement reaches
the first peak, the entanglement oscillates around $0.33$ when
$J_z=0.0$. When $J_z=0.5$, the entanglement oscillates around
$0.11$. For the above cases, the entanglement exists all the time.
While for $J_z=1$, the entanglement appears and disappears
alternatively. After $t=25$, the entanglement disappears all the
time. When $J_z=1.5, 2.0$, the entanglement disappears and never
appears after the first peak emergences. The entanglement decreases
as the anisotropic interaction increases. It is noted that the
propagation velocity of the entanglement is slightly influenced by
the anisotropy parameter, but the damping coefficient of the
entanglement wave is strongly dependent on the anisotropic parameter\cite{Amico,Amicoa}.
.

\section{Discussion}

By using the method of the adaptive time-dependent density-matrix
renormalization-group, the time evolution of the entanglement in a
one-dimensional spin-1/2 anisotropic XXZ model is investigated when
two quenches are performed. One quench is a sudden change of the
anisotropic interaction, while another is the abrupt release of the
maximum number of domain walls of staggered magnetization. The
dynamics of pairwise entanglement between the two nearest qubits in
the spin chain is studied. The entanglement of the two-spin qubits
can be created and oscillates after one of the two quenches is
performed. In both cases of the quench, the time evolution of the
entanglement is symmetric about the central qubit. For the
anisotropic interaction quench, an entanglement wall creates. Except
the first peak, the entanglement increases and oscillates as the
anisotropic interaction $J_z$ increases. For small $J_z$, the
entanglement may disappear at longer time. In the quench of domain
walls of staggered magnetization, the entanglement of central two
qubits appears first, after a short period, the entanglement of the
next two qubits appears, and so on. The height of the first peak in
the entanglement also decreases as the pair of two qubits moves away
from the center. The entanglement of the central two qubits
decreases when the anisotropic interaction increases, and then
increases when the anisotropic interaction increases further. For
two kinds of the quenches, the oscillation frequency of the
entanglement is strongly influenced by the anisotropy. This
phenomenon may be used to control the dynamics of the entanglement
by varying the anisotropic interaction of the Heisenberg spin chain.

\vskip 0.4 cm

\begin{acknowledgments}

It is a pleasure to thank Yinsheng Ling and Yinzhong Wu for their
many helpful discussions. The financial supports from the National
Natural Science Foundation of China (Grant No. 10774108).
\end{acknowledgments}


\clearpage
\newpage
\begin{figure}
\includegraphics[scale=0.5]{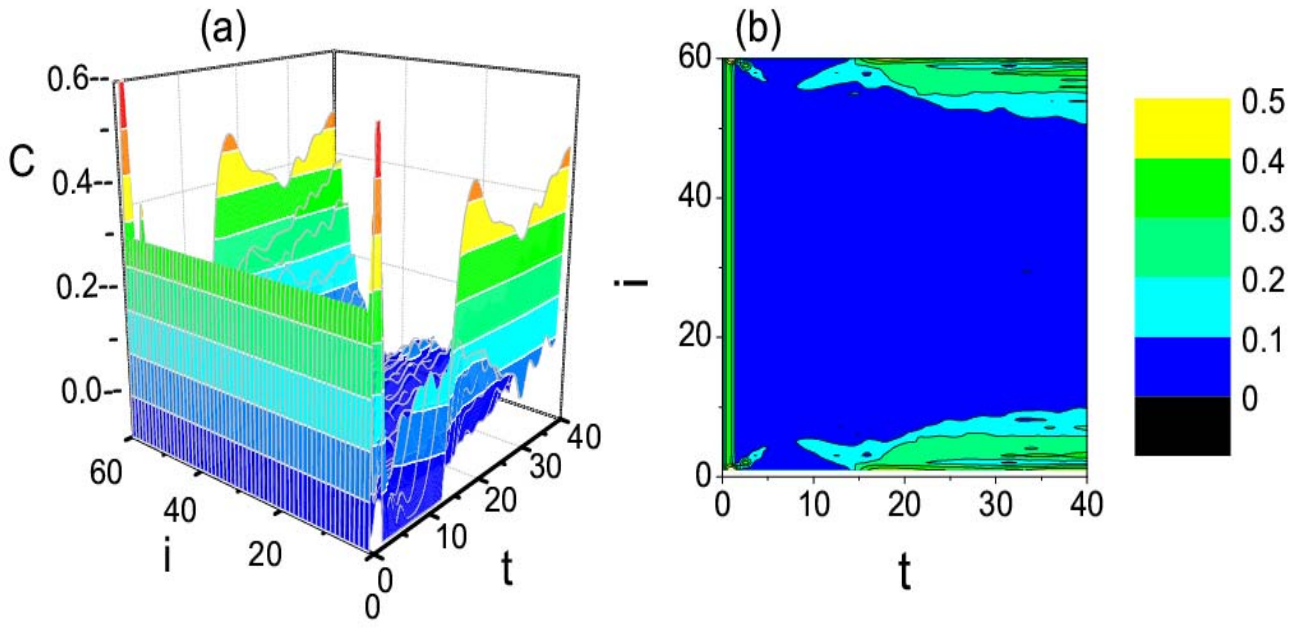}\caption{The pairwise entanglement $C_{i, i+1}$ between the two nearest
qubits is plotted as a function of spin site $i$ and time $t$ with
$J_z=1$ when there is anisotropic interaction quench. The size of
the system is $N=60$. (a). The three dimensional plot of
$C_{i,i+1}$. (b). The contour line of $C_{i,i+1}$.}
\end{figure}

\clearpage
\newpage
\begin{figure}
\includegraphics[scale=0.5]{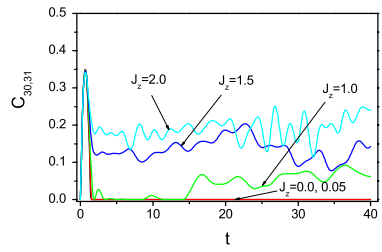}\caption{ The pairwise entanglement $C_{30, 31}$ of the two central qubits is
plotted as a function of time $t$ for different anisotropic
interaction $J_z$ when there is anisotropic interaction quench. The
sizes of the system is $N=60$.
}
\end{figure}

\clearpage
\newpage
\begin{figure}
\includegraphics[scale=0.5]{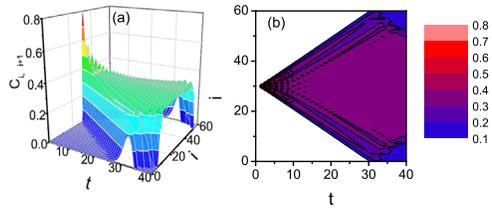}\caption{The pairwise entanglement $C_{i, i+1}$ between the two nearest
qubits is plotted as a function of spin site $i$ and time $t$ with
$J_z=1.0$ when there is domain walls quench. The size of the system
is $N=60$. (a). Three dimensional plot of $C_{i,i+1}$. (b). The
contour plot of $C_{i,i+1}$. }
\end{figure}

\clearpage
\newpage
\begin{figure}
\includegraphics[scale=0.5]{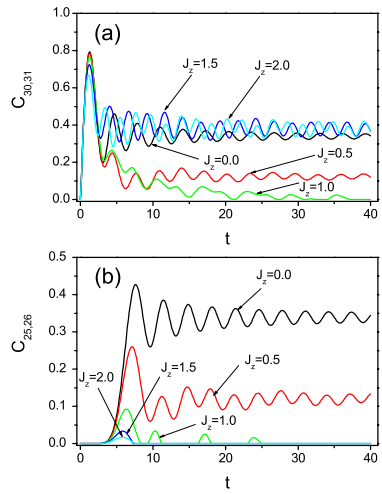}\caption{ The pairwise entanglement is plotted as a function of time $t$ for
different anisotropic interaction $J_z$ when there is domain walls
quench. The size of the system is $N=60$. (a). The pairwise
entanglement $C_{25, 26}$. (b). The pairwise entanglement $C_{30,
31}$ of the two central qubits. }
\end{figure}

\end{document}